# Effective chemical potential for non-equilibrium systems and its application to molecular beam epitaxy of Bi$_2$Se$_3$

Na Wang[1,2,3], Damien West[3], Xianran Xing[2], Wenhui Duan[1], and S. B. Zhang[3,4,*]

[1]Department of Physics, Tsinghua University, Beijing 100084, China

[2]Department of Physical Chemistry, University of Science and Technology Beijing, Beijing 100083, China

[3]Department of Physics, Applied Physics & Astronomy, Rensselaer Polytechnic Institute, Troy, NY 12180

[4]Beijing Computational Science Research Center, Beijing 100193, China

**Abstract**

First-principles studies often rely on the assumption of equilibrium, which can be a poor approximation, e.g., for growth. Here, an effective chemical potential ($\bar{\mu}$) method for non-equilibrium systems is developed. A salient feature of the theory is that it maintains the equilibrium limits as the correct limit. In application to molecular beam epitaxy, rate equations are solved for the concentrations of small clusters, which serve as feedstock for growth. We find that $\bar{\mu}$ is determined by the most probable, rather than by the lowest-energy, cluster. In the case of Bi$_2$Se$_3$, $\bar{\mu}$ is found to be highly supersaturated, leading to a high nucleus concentration in agreement with experiment.

*zhangs9@rpi.edu

The concept of chemical potential is one of the most fundamental quantities associated with thermodynamic equilibrium. It generalizes the effect of the environment on the energetics of a single type of particle. Through the use of atomic chemical potentials, we can compare the energies of systems consisting of different numbers of atoms. In defect physics, this plays a central role in the calculation of the formation energies of defects and allows for a statistical determination of defect concentrations which determine the electronic properties of semiconductors. Further, surface reconstruction and even the shape of nanocrystals are intimately tied with the availability of different chemical species, with the surface energy density generally depending on the atomic chemical potentials of the various species. While equilibrium theory is almost exclusively used to determine the limits of the atomic chemical potentials in the calculation of energetics of different structures under different growth conditions, this is an outstanding contradiction in the theoretical community as growth itself is a highly non-equilibrium process.

The chemical potential, $\mu$, is a thermodynamic quantity, defined by $\mu = \frac{\partial G}{\partial n}$, where $G$ is the Gibbs free energy and $n$ is the number of the $n^{\text{th}}$ species [1]. For equilibrium or near-equilibrium growth (NEG), there is a well-defined relationship between the $\mu'$s of the species in the system [1]. Consider, for example, a $A_m B_n$ binary, we have

$$m\mu_A + n\mu_B = \mu_{A_m B_n}. \tag{1}$$

As a result of Eq. (1), only one $\mu$ is an independent variable, i.e., one can use $\mu_B$ for this purpose and $\mu_A$ will be given by $\mu_A = (\mu_{A_m B_n} - n\mu_B)/m$. Should $\mu$ for a species $i$ exceeds its upper bound $\mu_{i0}$, defined as the chemical potential of the corresponding elemental solid or gas, this species will precipitate [2,3], thereby preventing any further increase of $\mu$. In other words,

$$\mu_i \leq \mu_{i0} \ (i = A, B). \tag{2}$$

Using the $\mu'$s, the state of the system, whether it is a defect, a surface, or a crystal structure (denoted here as X) is determined by its formation energy with respect to bulk,

$$\Delta H_f[X] = E_{tot}[X] - E_{tot}[\text{bulk}] - \sum n_i \mu_i, \tag{3}$$

where $E_{tot}$ is the total energy [3].

This NEG theory has been applied to numerous materials behaviors with ample successes, ranging from surface reconstruction [2,4] and defect physics [3,5], to nanostructures [6,7]. For systems in which a subsystem of concern can be isolated from the rest due, for example, to high reaction barriers, one can also use the NEG theory, provided that one can establish an approximate equilibrium within the subsystem [8-10]. However, when such a division is not obvious and the system is far away from equilibrium, the NEG theory can run into problems. For instance, recent interest in the physics of topological insulators (TIs) has spurred considerable activity to grow high-quality TIs such as $Bi_2Se_3$ with controlled properties [11-16]. Molecular beam epitaxy (MBE) is widely accepted as an effective method to grow high-quality films [13-16]. However, the defect density is still far from being satisfactory, due in part to the small size of the islands. Strictly speaking, MBE is not an equilibrium process [17]; its outcome may rely heavily on the dynamics of the source atoms/molecules. On the other hand, MBE is relatively simple for its slow growth rate and the simplicity of the sources. This raises the question: can one incorporate kinetics into the chemical potential-based approach for non-equilibrium process?

Note that considerable theoretical efforts have been made in the past to study growth. Analytical rate equation approach, based on empirical kinetic parameters, has been employed to study island formation and growth [18-24]. Statistical approaches such as kinetic Monte-Carlo simulation and molecular dynamics have been used to study film growth [25-30]. Combined with the above approaches, first-principles density functional theory (DFT) calculations have also been attempted. It appears that these non-equilibrium approaches do not use the concept of chemical potential, which had served as a basis for NEG, except perhaps for the thermodynamic theory of nucleation [31]. Recently, a renewed interest in the chemical potential has emerged, as it can help choosing lowest-energy paths for growth [32].

In this paper, we derive a generic effective chemical potential ($\bar{\mu}$) theory, which is suited for non-equilibrium physics. Taking the MBE growth of $Bi_2Se_3$ as an example, the use of $\bar{\mu}$ allows us to simplify the overall complex process. We divide the non-equilibrium growth (nonEG) into three stages: pre-nucleation, nucleation, and island growth. For pre-nucleation, we determine $\bar{\mu}$ by explicitly solving rate equations. First-principles calculation shows that $\bar{\mu}$ is solely determined by the most probable, rather than by the lowest-energy, cluster(s) on the surface. For nucleation, it is found that the nucleation barrier vanishes when the critical size of the clusters is only a few molecule large due to the highly supersaturated nature of $\bar{\mu}$. This results in a high concentration of nuclei – a conclusion that contradicts the NEG model, but is in qualitative agreement with experiment.

One can define chemical potentials for an individual cluster $\alpha$, irrespective of equilibrium [1,33],

$$\mu_\alpha = \frac{\partial G}{\partial c_\alpha} = \frac{\partial(\sum_\beta G_\beta)}{\partial c_\alpha} = \frac{\partial G_\alpha}{\partial c_\alpha} = E_{0\alpha} - kT \ln \frac{N_S}{c_\alpha}, \quad (4)$$

where $c_\alpha$ is the areal concentration, $G_\alpha$ is the Gibbs free energy, $E_{0\alpha}$ is the energy of the cluster on the surface relative to the isolated constituent atoms, $N_S$ is the number of sites per area that can hold the cluster (which, for simplicity, has been assumed to be much larger than the number of clusters), and $k$ and $T$ have the standard definitions. Here, considering a binary AB system ($\alpha = A_s B_t$) as in NEG, we postulate that $\bar{\mu}$ that enters Eq. (3) is given by

$$\bar{\mu}_A \left[ \sum_s s \left( \sum_t c_{A_s B_t} \right) \right] + \bar{\mu}_B \left[ \sum_t t \left( \sum_s c_{A_s B_t} \right) \right] = \sum_{s,t} c_{A_s B_t} \mu_{A_s B_t}, \quad (5)$$

where the sums over $s$ and $t$ run over all possible AB clusters. As the concentrations of the clusters depend on the kinetics of the system, this weighted average incorporates both the energetics and kinetics in a simple way. When the system contains only a single element, Eq. (5) is reduced to

$$\bar{\mu}_{A0} = \frac{\sum_s c_{A_s} \mu_{A_s 0}}{\sum_s s c_{A_s}}, \quad (6)$$

where the subscript 0 denotes single-element quantity, as in Eq. (2). For a binary system, as long as the formation of $A_s B_t$ from A and B clusters is exothermic (*i.e.*, with $\Delta E < 0$), Eq. (6) cannot hold except when species A (or B) starts to precipitate as pure clusters. Hence, it sets an upper bound for $\bar{\mu}_A$. The same is true for species B. Therefore,

$$\bar{\mu}_i \leq \bar{\mu}_{i0} \ (i = A, B). \quad (7)$$

It is necessary to point out that Eqs. (5) and (7) are analogous to Eqs. (1) and (2), respectively.

To calculate $\bar{\mu}$, we solve the following rate equations [31],

$$\dot{c}_\alpha = F_\alpha - k_\alpha^{des} c_\alpha + \sum_i \left( k_{\alpha-i}^{+i} c_{\alpha-i} c_i + k_{\alpha+i}^{-i} c_{\alpha+i} - k_\alpha^{-i} c_\alpha - k_\alpha^{+i} c_\alpha c_i \right), \quad (8)$$

$$k_\alpha^{des} = \nu \, \mathrm{Exp}(-\varphi_\alpha^{des}/kT), \quad (9)$$

$$k_\alpha^{-i} = \nu \, \text{Exp}(-\varphi_\alpha^{-i}/kT), \quad (10)$$

$$k_\alpha^{+i} = 2\pi(D_\alpha + D_i), \quad (11)$$

$$D_\alpha = a^2 \nu \, \text{Exp}(-\varphi_\alpha^{diff}/kT), \quad (12)$$

where $F_\alpha$ is the deposition rate, $k_\alpha^{des}$ is the desorption rate, $k_\alpha^{-i}$ and $k_\alpha^{+i}$ are the dissociation and association rates by emitting and absorbing a cluster $i$, respectively, $D_\alpha$ is the diffusion coefficient [34] with $a$ being the in-plane lattice constant, and $\nu \sim 10^{13} s^{-1}$ is the vibrational frequency. The use of Eq. (11) above assumes that the process is diffusion limited [See discussion in Supplementary Information (SI) and Fig. S1 for the general case]. For simplicity, in the following we consider only a flat surface, so the effect of the Ehrlich-Schwoebel barrier [31] can be ignored. We also assume a steady-state, *i.e.*, letting $\dot{c}_\alpha \equiv 0$ in Eq. (8).

Note that in the limit when all the barriers in Eqs. (8)-(12) can be ignored, the system is dominated by its lowest-energy form, namely, the concentrations for all clusters, as well as for all finite-sized islands diminish, leaving only the $A_m B_n$ bulk. In this case, Eqs. (5) and (7) become Eqs. (1) and (2) respectively, *i.e.*, the system approaches its NEG limit. Also note that chemical potential is not an easily measurable quantity. People may, instead, use the off-stoichiometry of the system, which can be directly measured, as an indirect indicator of $\mu_i$. This is true not only for the NEG model [35] but also for the nonEG model here.

First-principles calculations were performed using the VASP code [36] to determine the relevant cluster size, structure, and energy barrier. The projector augmented wave (PAW) method [37] and the local density approximation (LDA) to the exchange-correlation functional [38] were used. Plane waves with a kinetic energy cutoff of 210 eV were used as the basis set. Integration of the Brillouin zone was done with sufficient k-point sampling [39] such that the numerical error is less than 0.01 eV. All atoms were fully relaxed until forces on the atoms were less than 0.01 eV/Å. The improved tangent finding method [40], within the framework of the nudged elastic band (NEB) method, was used to determine the energy barriers. Experimental growth temperature of 500 K [13-15] was assumed in the rate calculations.

Denoting $t_0$ as the time when nucleation takes place, the aforementioned 3 growth stages can be restated as follows: stage 1: $t \leq t_0$ (pre-nucleation), stage 2: $t \sim t_0$ (nucleation), stage 3:

$t \gg t_0$ (island growth). Figure 1 illustrates schematically the evolution of the size of the clusters, and those of the nuclei and islands, and the corresponding $\mu$'s.

In stage 1, there are only clusters, no nuclei, on the surface of $Bi_2Se_3$, which in the calculation is taken as five atomic layers, or one quintuple layer (QL). Because the binding between QLs is van der Waals (vdW) [11-13], adding QLs to the substrate is unlikely to affect our results. Figure 2 depicts the clusters we considered in our ab-initio calculations. They are organized according to the chart in Fig. 2(a), where dashed line denotes the cutoff size of the clusters considered in this study (see details in Fig. S2, SI). Our calculations show that stable clusters on the surface are the same as those in vacuum. This can be expected as Bi tends to form 3 bonds, whereas Se tends to form 2 bonds, in line with the octet rule [41]. Figure 3 shows the calculated diffusion barrier ($\varphi_\alpha^{diff}$), desorption barrier ($\varphi_\alpha^{des}$), and dissociation barrier ($\varphi_\alpha^{-i}$) for the clusters: $\varphi_\alpha^{diff}$ is generally small, often a couple tenth of an eV, except for atomic Se and $BiSe_2$. $\varphi_\alpha^{des}$ is higher, especially for smaller clusters. $\varphi_\alpha^{-i}$ is also generally higher and depends sensitively on the reaction pathway. Both $\varphi_\alpha^{des}$ and $\varphi_\alpha^{-i}$ exhibit a large variation over a couple eV.

Figures 4(a-b) show $c_\alpha$ and the corresponding $\mu_\alpha$, which reveal that the highest concentrations of clusters are that of atomic Se and $BiSe_2$, respectively. At first glance, it may appear counterintuitive that some clusters with lower $\mu_\alpha$, e.g., $Bi_2Se_2$ and $Bi_2Se_3$, also have low $c_\alpha$. This is because of the low-desorption barrier of $Bi_2Se_3$ in Fig. 3, which depletes not only $Bi_2Se_3$ but also $Bi_2Se_2$ (via an association with atomic Se, see Fig. S1). In MBE, the Se partial pressure is much higher than that of Bi. As a result, the population of Se is determined almost entirely by *pure* Se clusters via Eq. (7),

$$\bar{\mu}_{Se} = \bar{\mu}_{Se0} = \mu_{Se}(\text{atom}). \tag{13}$$

And for Bi, one has

$$\bar{\mu}_{Bi} = \frac{\sum_{s,t} c_{Bi_sSe_t} \mu_{Bi_sSe_t} - \mu_{Se}(\text{atom})[\sum_t t(\sum_s c_{Bi_sSe_t})]}{[\sum_s s(\sum_t c_{Bi_sSe_t})]}. \tag{14}$$

These $\bar{\mu}$ values, $\bar{\mu}_{Bi} = -4.44$ eV and $\bar{\mu}_{Se} = -3.82$ eV, are given in Fig. 4(b) as horizontal dashed lines. Our study reveals that it is *the most-probable, not the lowest-energy,* clusters that determine $\bar{\mu}$. This conclusion is in line with the basic principles of statistical physics, irrespective of the computational details, whereby it reinforces the notion that using $c_\alpha$ as the weighting factor for $\bar{\mu}$ in Eq. (5) is physically correct.

Note that there is a large (3.08 eV) difference between the current model ($2\bar{\mu}_{Bi} + 3\bar{\mu}_{Se} = -20.34$ eV) and the NEG model ($2\mu_{Bi} + 3\mu_{Se} = -23.42$ eV). The fact that the former is significantly higher than the latter gives rise to supersaturation during the growth. In other words, Eq, (7) sets a new set of boundaries for chemical potentials as the prime reason that account for the differences between nonEG and NEG. Importantly, this finding allows one to reexamine defects in $Bi_2Se_3$, as it suggests that native defects may be more easily formed than predicted by NEG theory [42], highlighting experimental challenges in growing high-quality epitaxial films [43].

Before finishing up the discussion of stage 1, we would like to point out that our consideration of the availability of clusters is important, but only a first, step in improving upon NEG theory. Within the framework of the current development, higher order effects (such as the kinetics associated with the incorporation of available species into the growth front) can also be incorporated, which however will be differed to future work. For stages 2 and 3, i.e., nucleation and island growth, due to continued deposition, we do not expect the concentrations of the small clusters to change considerably (as will be shown later).

In stage 2, owing to the formation of nuclei, bifurcation of $\mu$ will take place – one for the nuclei ($\mu_{nuc}$) and one for the small clusters ($\bar{\mu}$), as depicted in Fig. 1(b). Bifurcation reflects the fact that once nuclei are formed, it is energetically favorable for them to grow further by consuming available smaller clusters on the surface, rather than consuming themselves. This should be contrasted to NEG where bifurcation of $\mu$ is strictly forbidden. Once $\bar{\mu}$ for clusters are determined, one may approach the nucleation problem in different ways. For example, to study the temperature T dependence of nucleus density, one could apply Eqs. (4)-(12) to potential nuclei [which should be among the larger clusters in Fig. 2 and beyond (= embryo nuclei)] to determine $\mu_{nuc}$ and $c_{\alpha,nuc}$, respectively. The non-negligible association barriers [see, e.g., Fig. S1(a) for clusters] suggest that increasing T could help increasing the nucleus size and hence decreasing its density. However, the window for the increase can be limited, as although dissociation barriers are usually higher than association barriers [e.g., by comparing Fig. S1(a) with Fig. 3(c)], the difference may not be that dramatic. As a result, further increasing T could also lead to the dissociation of already-formed nuclei, thereby increasing the nucleus density. Although rather qualitative, the conclusion here is already in line with experiments [14,44].

If, however, our interest is only on the critical size of the nuclei, not on its T dependence, explicitly invoking kinetic theory here can be an overkill. Instead, we only need calculate the nucleation barrier, which is defined as the maximum of the Gibbs free-energy [31],

$$\Delta G = \Delta E_{\text{tot}} - (n_{\text{Se}}\bar{\mu}_{\text{Se}} - n_{\text{Bi}}\bar{\mu}_{\text{Bi}}). \tag{15}$$

After reaching the critical size, adding additional atoms/clusters to the nuclei is energetically favored. Using $\bar{\mu}$ in stage 1, we obtain Fig. 5, showing that $\Delta G$ is remarkably small and even negative for $(Bi_2Se_3)_n$ clusters with $n > 3$. Usually, $\Delta G$ for cluster of these sizes is significantly positive due to its high surface energy. Due to high $\bar{\mu}$ and the relatively low energy of the vdW surface, however, this is no longer the case here. Figure S3 shows that except for $n = 1$ and 3, the desorption barriers are reasonably large so the nuclei in Fig. 5 do grow into islands. The small nuclei size is consistent with measured high island concentration, $c_{isl} \sim 10^9 \text{-} 10^{10}$ cm$^{-2}$ [14,15].

In stage 3, the basic physics should be the same as in stage 2. Due to the presence of the islands, however, one should recalculate $c_\alpha$ and $\bar{\mu}$ by adding a term $-k_{isl}^{+i} c_{isl} c_i$ to Eq. (8), where $c_{isl}$ is the concentration of the islands and $k_{isl}^{+i} = 2\pi(D_i)$. Figure 4(c) shows the results for a typical $c_{isl} = 10^9$ cm$^{-2}$. More results on the $c_{isl}$-dependence can be found in Fig. S4, SI. As it turns out, neither $c_\alpha$ nor $\bar{\mu}$ is changed significantly. As a comparison, Fig. 4(c) also shows $c_\alpha$ under the NEG condition. They are many orders of magnitude smaller.

In summary, an effective chemical potential approach for non-equilibrium growth is developed. We find that $\bar{\mu}$ for an atomic species is determined primarily by the most probable cluster in which the atom resides during growth. Because "most probable" is a *balance* between energetics and kinetics, our findings thus set a new criterion for most relevant events during growth. Application to MBE growth of Bi$_2$Se$_3$ suggests that $\bar{\mu}$ is highly supersaturated, resulting in an exceedingly small critical size of nuclei. While a high concentration of islands is in agreement with experiment, out results also reveal that to grow better-quality films requires finding ways to stabilize the most probable clusters, thereby lowering $\bar{\mu}$. Our formulation is general. It may be used to study structure, surface morphology, and shape of a nanoparticle, as well as defect and impurity, in non-equilibrium-grown solids. While kinetic theory has been around for a long time, most first-principles studies today are still based on the NEG model. Our chemical potential-based development here offers a natural vehicle to transform such bulk studies into the more experimentally relevant non-equilibrium regime.

We thank X. Liu and J. Wang for stimulating discussions. Chemical potential theory was supported by the US Department of Energy (DOE) under Grant No. DESC0002623. $Bi_2Se_3$ growth study was supported by NSF Award No. EFMA-1542798. Work in China was supported by the Ministry of Science and Technology of China (Grant Nos. 2011CB921901 and 2011CB606405), and the National Natural Science Foundation of China (Grant No. 11334006). Supercomputer time was provided by NERSC under the Grant No. DE-AC02-05CH11231 and the Center for Computational Innovations (CCI) at RPI.

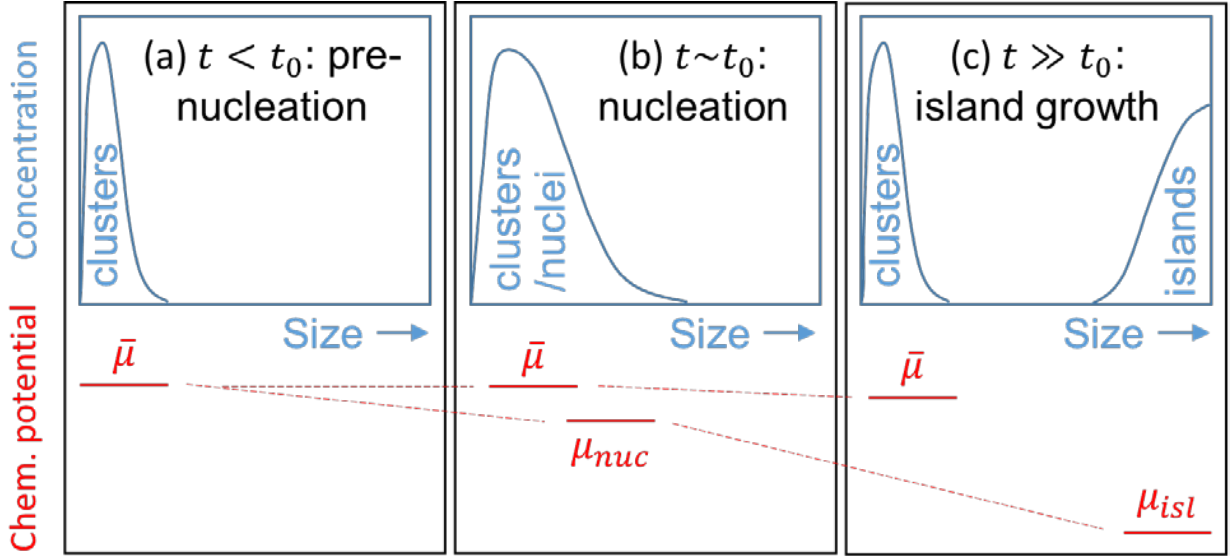

FIG. 1: (Color online) A schematic illustration of cluster/island distribution and the corresponding chemical potentials in three different stages of $Bi_2Se_3$ growth: (a) pre-nucleation, (b) nucleation, and (c) island growth. $t_0$ is the characteristic nucleation starting time.

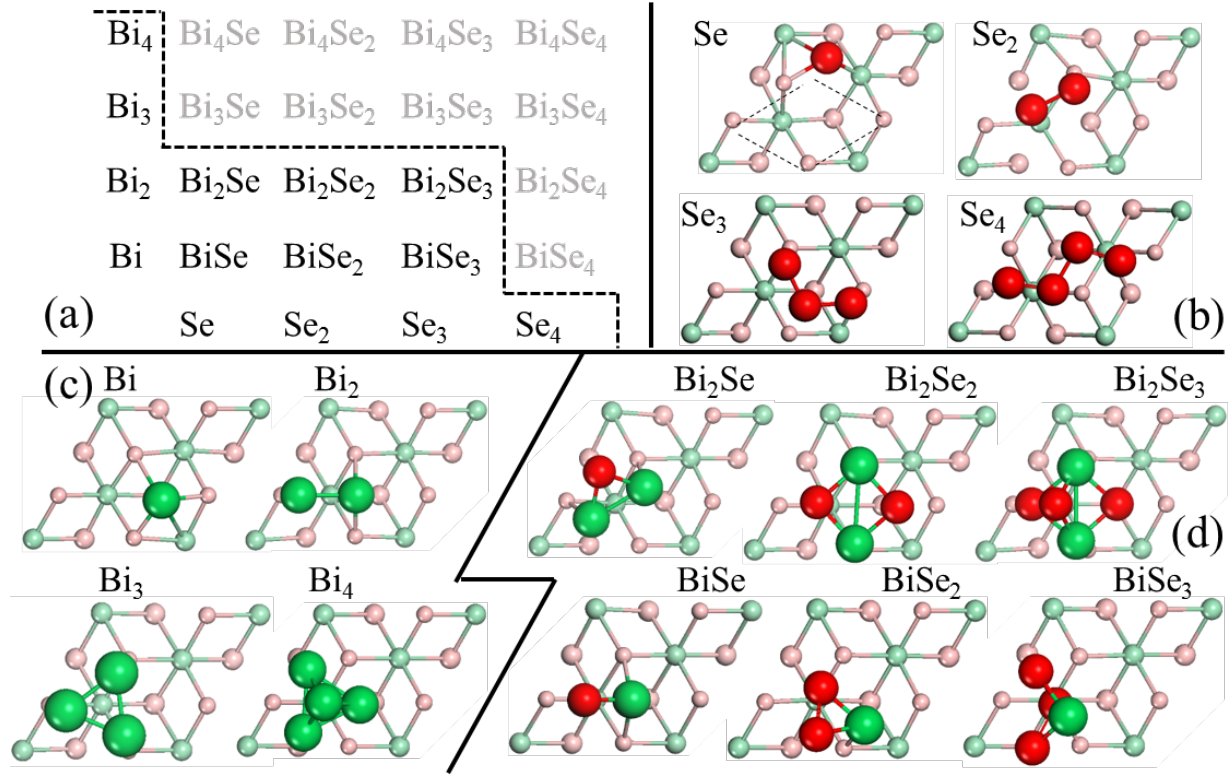

FIG. 2: (Color online) (a) Small clusters for $Bi_2Se_3$ growth. They are organized in terms of size: horizontal = increasing number of Se atoms; vertical = increasing number of Bi atoms. Dashed zigzag line is discussed in the main text. (b)-(d) Calculated most-stable cluster structures on a $Bi_2Se_3$ (0001) substrate: (b) pure Se, (c) pure Bi, and (d) mixed $Bi_sSe_t$ clusters. Red atoms are Se whereas green atoms are Bi.

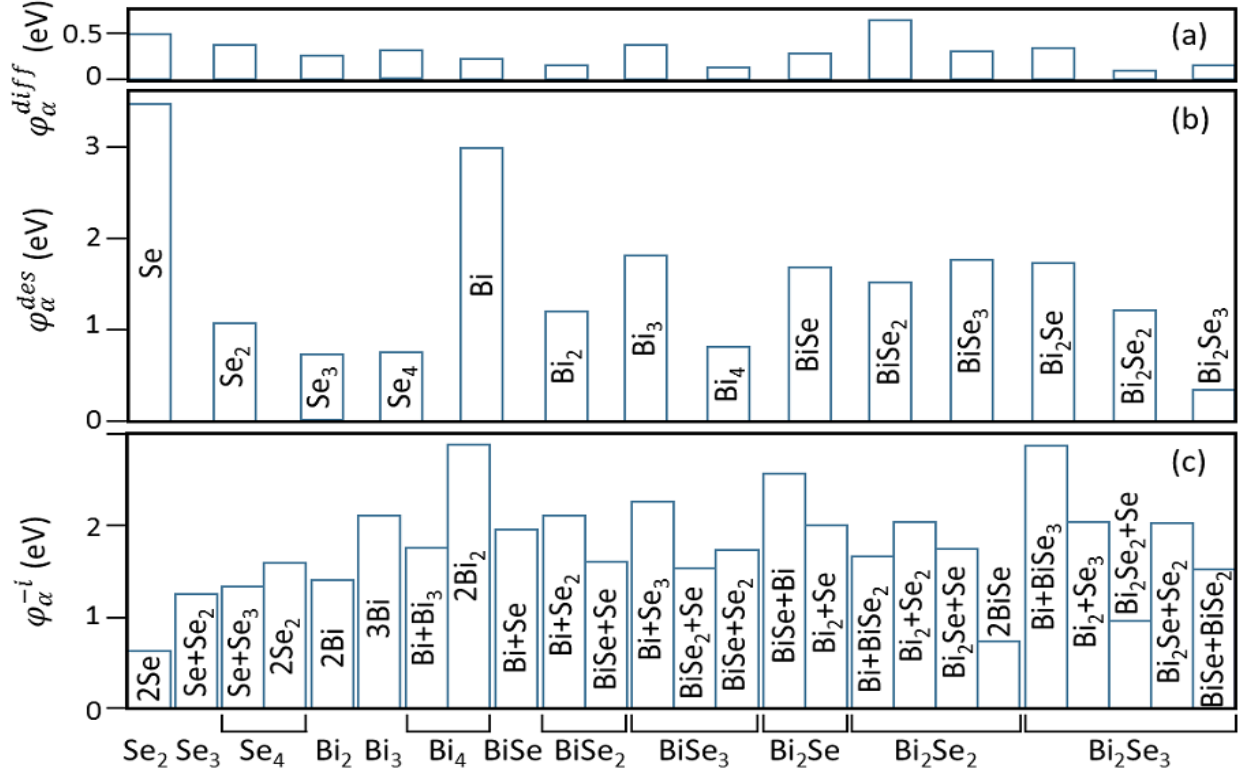

FIG. 3: (Color online) Calculated barriers for (a) diffusion ($\varphi_\alpha^{diff}$), (b) desorption ($\varphi_\alpha^{des}$), and (c) dissociation ($\varphi_\alpha^{-i}$) of the clusters in Fig. 2. Cluster names for both (a) and (b) are given in (b). For cluster dissociation in (c), a cluster labeled in the horizontal axis can have multi-reaction pathways, which are labeled inside or by the vertical bars.

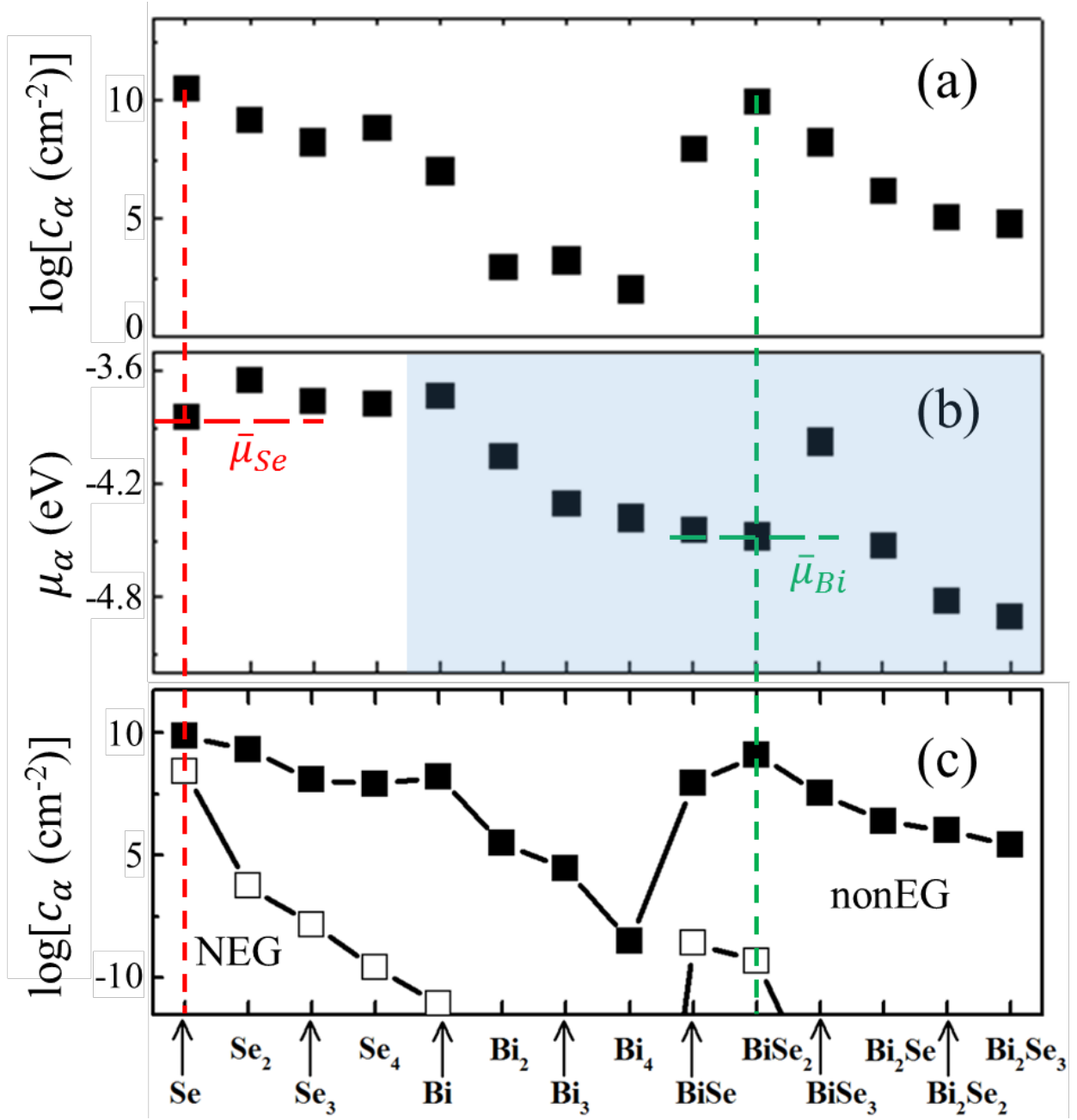

FIG. 4: (Color online) (a) Logarithmic concentration of small clusters in stage 1. (b) The corresponding chemical potentials by Eq. (4) and $\bar{\mu}$'s by Eqs. (13) and (14), respectively. (c) Logarithmic concentration of small clusters in stage 3. Solid squares are based on the nonEG model with $c_{island} = 10^9 \text{cm}^{-2}$, whereas open squares are the prediction by the NEG model.

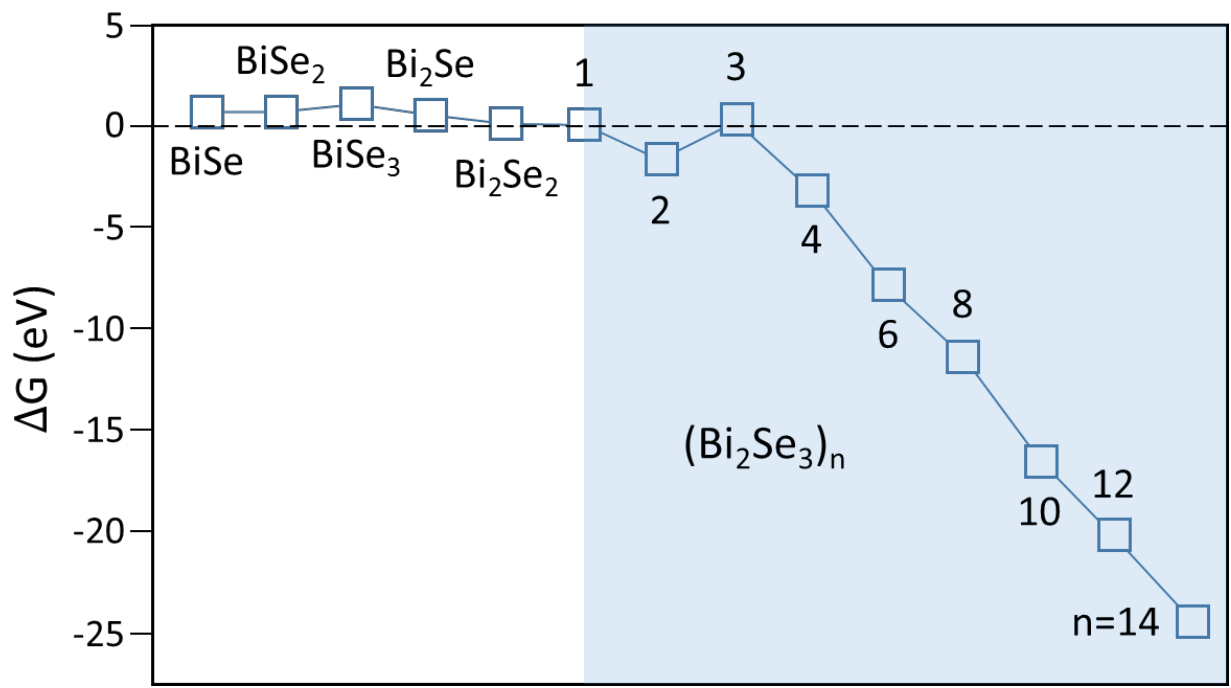

FIG. 5: (Color online) Free-energy barrier $\Delta G$ for nucleation on various clusters. Shaded area is where only stoichiometric molecular clusters $(Bi_2Se_3)_n$ are considered in the calculation.